\documentclass[journal=jacsat,manuscript=article]{achemso}

\usepackage[version=3]{mhchem} 
\usepackage{amsmath,amssymb,mathrsfs,accents}
\usepackage{color}
\usepackage[scr=rsfso]{mathalpha}
\usepackage{booktabs}                   
\usepackage{multirow}                   
\usepackage{diagbox}                    
\usepackage{bm}
\usepackage{physics}
\usepackage[T1]{fontenc}
\usepackage[utf8x]{inputenc}

\author{Yu-Chen Wei}
\affiliation{Institute of Atomic and Molecular Sciences, Academia Sinica, Taipei, Taiwan}
\alsoaffiliation{Department of Chemistry, National Taiwan University, Taipei, Taiwan}
\author{Liang-Yan Hsu}
\affiliation{Institute of Atomic and Molecular Sciences, Academia Sinica, Taipei, Taiwan}
\alsoaffiliation{Department of Chemistry, National Taiwan University, Taipei, Taiwan}
\alsoaffiliation{National Center for Theoretical Sciences, Taipei, Taiwan}
\email{lyhsu@gate.sinica.edu.tw}

\title[An \textsf{achemso} demo]
  {Polaritonic Huang-Rhys Factor: Basic Concepts and Quantifying Light-Matter Interaction in Medium
}

\begin{document}







\begin{abstract}
   Huang-Rhys (HR) factor, a dimensionless factor that characterizes electron-phonon coupling, has been extensively employed to investigate material properties in various fields. In the same spirit, we present a quantity called polaritonic HR factor to quantitatively describe the effects of (i) light-matter coupling induced by permanent dipoles and (ii) dipole self-energy. The former can be viewed as polaritonic displacements, while the latter is associated with the electronic coupling shift. In the framework of macroscopic quantum electrodynamics, the polaritonic HR factor, coupling shift, and modified light-matter coupling strength in an arbitrary dielectric environment can be evaluated without free parameters, whose magnitudes are in good agreement with the previous experimental results. In addition, polaritonic progression developed in our theory indicates that large polaritonic HR factors can result in light-matter decoupling, multipolariton formation, and non-radiative transition. We believe that this study provides a useful perspective to understand and quantify light-matter interaction in medium.
\end{abstract}

Photons, quanta of the electromagnetic (EM) field, are powerful detectors used to probe optoelectronic properties of molecules, assuming that incident electromagnetic fields during measurements do not affect molecular intrinsic properties. Recently, several studies have demonstrated that the confinements of EM fields can alter potential energy surfaces and chemical reaction dynamics through strong coupling between vacuum EM fluctuations and molecules \cite{thomas2019tilting,nagarajan2021chemistry,ebbesen2016hybrid,garcia2021manipulating}. These results open a new era in exploring the fundamental principles of chemistry based on quantum electrodynamics (QED) and draw considerable attention to investigate the mechanisms of chemical reactions driven by strong light-matter interactions \cite{herrera2016cavity,flick2017atoms,schafer2019modification,pavovsevic2022cavity,garcia2021manipulating,yuen2019polariton,ribeiro2018polariton,li2020cavity,li2022molecular}. 

Light-matter couplings induced by permanent dipoles and dipole self-energy  have been studied for a long time \cite{meath1984effects,bavli1991nonlinear,barut1987quantum,meath1984importance,bavli1990nonlinear},
which are considered to be key factors for modifying chemical reactions through strong light-matter couplings \cite{flick2017atoms,rokaj2018light,frisk2019ultrastrong,george2016multiple,schafer2020relevance,semenov2019electron,mandal2020polarized}. For instances, by employing cavity QED, Semenov and Nitzan derived an electrodynamic Franck-Condon-type factor associated with permanent dipoles and dipole self-energy. The results show that large electrodynamic Franck-Condon-type factor suppresses electron transfer rates in the normal region \cite{semenov2019electron}. In addition, Mandal et~al. considered the non-relativistic Pauli–Fierz Hamiltonian and demonstrated that permanent dipoles and dipole self-energy can lead to the generation of multiple photons from a single exothermic electronic excitation, effectively achieving the dynamical Casimir effect \cite{mandal2020polarized}. Furthermore, Sch{\"a}fer et~al. validated that discarding self-interaction terms would cause unphysical artifacts including loss of gauge invariance, basis set dependence, loss of bound states, formation of radiating ground states, and artificial dependence on the static dipole in nonrelativistic quantum electrodynamics (QED) \cite{schafer2020relevance}. 
 
 The QED effects of permanent dipoles and dipole self-energy play critical roles in strong light-matter coupling regime, but how to quantitatively estimate these factors are still lacking. The lack of quantitative estimates of light-matter interactions may result in inappropriate interpretations in experiments and further impede quantitative prediction of QED-based chemical and physical processes. To solve the above issues, we present a new physical quantity, polaritonic Huang-Rhys (HR) factor, based on macroscopic QED in order to quantify light-matter coupling strength caused by dipole self-energy and permanent dipoles. In addition, we propose several useful concepts, including polaritonic displacements, reorgnization dipole-self coupling and polaritonic progression by analogy with the physical picture of vibronic interaction. By virtue of macroscopic QED, polaritonic HR factors can be evaluated in various dielectric environments without free parameters. We believe that the concept of the polaritonic HR factor and its corresponding theory based on macroscopic QED will enable intuitive understanding and quantitative analysis of intriguing phenomena in the strong light-matter coupling regime.

 To properly describe the light-matter interaction in a dispersive and absorbing dielectric environment, we start from the minimal coupling (i.e., spinless) Hamiltonian in the framework of macrospcopic QED \cite{gruner1996green,dung1998three,buhmann2013dispersion}. After making the long-wavelength approximation and the G{\"o}eppert-Mayer transformation, one can derive the corresponding length-gauge Hamiltonian as follows (see sections 1 and 2 in the Supporting Information (SI)),
\begin{align}
\label{Eq:mininal coupling_H_GM}
\hat{H}_\mathrm{GM}=&~\sum_\alpha\frac{\hat{\mathbf{p}}_\alpha^2}{2m_\alpha}+\frac{\mathrm{1}}{2}\int d\mathbf{r}\hat{\rho}_{\mathrm{M}}(\mathbf{r})\cdot\hat{\phi}_{\mathrm{M}}(\mathbf{r})\nonumber\\
&\hspace{0.3cm}+\int d\mathbf{r}\int_0^\infty d\omega\hbar\omega \hat{\mathbf{f}}^\dag(\mathbf{r},\omega)\cdot\hat{\mathbf{f}}(\mathbf{r},\omega)-\hat{\boldsymbol{\mu}}\cdot\hat{\mathbf{E}}(\mathbf{r}_\mathrm{M})+\frac{|\hat{\boldsymbol{\mu}}|^2}{2\epsilon_0 V_\mathrm{eff}},
\end{align}
where $m_\alpha$ and $\hat{\mathbf{p}}_\alpha$ stand for the mass and the momentum operator of the $\alpha^\mathrm{th}$ particle, respectively. The symbol $\hat{\rho}_{\mathrm{M}}(\mathbf{r})$ denotes a charge density operator for the molecule. The symbols $\hat{\phi}_{\mathrm{M}}(\mathbf{r})$ and $\hat{\phi}(\mathbf{r})$ correspond to molecular and external Coulomb potential operators, respectively. $\hat{\mathbf{f}}^\dag(\mathbf{r},\omega)$ ($\hat{\mathbf{f}}(\mathbf{r},\omega)$) is the creation (annihilation) operator for bosonic vector fields (polariton) in macroscopic QED.\cite{gruner1996green,dung1998three,buhmann2013dispersion} Note that the polaritons here represent photons dressed by medium polarization, which do not cover charged particles in the molecular Hamiltonian \cite{gruner1996green,dung1998three,buhmann2013dispersion,wei2021can}. The term $\hat{\boldsymbol{\mu}}\cdot\hat{\mathbf{E}}(\mathbf{r}_\mathrm{M})$ describes the light−matter interaction in the electric-dipole form, where $\hat{\boldsymbol{\mu}}$ and $\hat{\mathbf{E}}(\mathbf{r}_\mathrm{M})$ represent the dipole moment operator and the electric field operator, respectively. The last term is the
dipole self-energy, where $V_\mathrm{eff}$ is the effective mode volume of the transverse EM fields. Note that the dipole-self energy arises from the polarization of the molecule acting back on transverse EM fields \cite{cohen1997photons,mandal2020polarized} and can be expressed in terms of the inverse of transverse delta function \cite{gruner1996green,dung1998three,buhmann2013dispersion}, which is associated with $V_\mathrm{eff}$ (see SI Sections 1 and 2).

According to the previous studies,\cite{craig1998molecular,mandal2020polarized,chuang2022tavis} the first two terms in Eq.~\ref{Eq:mininal coupling_H_GM} correspond to the standard molecular Hamiltonian. As a result, these two terms can be simplified as a two-level system in the electronic subspace. In this framework, we regroup the terms in Eq.~\ref{Eq:mininal coupling_H_GM} as $\hat{H}_\mathrm{ele}-$the electronic Hamiltonian,  $\hat{H}_\mathrm{pol}-$the polaritonic Hamiltonian and $\hat{H}_\mathrm{mol\scalebox{0.7}{-}pol}-$ the light-molecule interaction as follows.
\begin{align}
\label{Eq:total_H}
    &\hat{H}_\mathrm{GM}=\hat{H}_\mathrm{mat}+\hat{H}_\mathrm{pol}+\hat{H}_\mathrm{mat\scalebox{0.7}{-}pol},\\
\label{Eq:ele_H}
    &\hat{H}_\mathrm{mat}=\epsilon_{\mathrm{g}}\ket{\mathrm{g}}\bra{\mathrm{g}}+(\hbar\omega_\mathrm{eg}+\epsilon_\mathrm{g})\ket{\mathrm{e}}\bra{\mathrm{e}},\\
\label{Eq:pol_H}
    &\hat{H}_\mathrm{pol}=\int d\mathbf{r}\int_0^\infty d\omega\hbar\omega \hat{\mathbf{f}}^\dag(\mathbf{r},\omega)\cdot\hat{\mathbf{f}}(\mathbf{r},\omega),\\
\label{Eq:mol_pol_H}
    &\hat{H}_\mathrm{mol\scalebox{0.7}{-}pol}=-\hat{\boldsymbol{\mu}}\cdot\hat{\mathbf{E}}(\mathbf{r}_\mathrm{M})+\frac{|\hat{\boldsymbol{\mu}}|^2}{2\epsilon_0 V_\mathrm{eff}}.
\end{align}
Here the symbols $\mathrm{e}$ and $\mathrm{g}$ denote electronic excited and ground state, respectively. The electronic ground state energy of the molecule $\epsilon_{\mathrm{g}}$ is set as zero, and the frequency $\omega_\mathrm{eg}$ stands for the electronic transition frequency between the electronic ground and excited states. Notably, the vibrational degrees of freedom are neglected in the present study since we would like to emphasize the interaction between the electronic states of a molecule and EM fields. In fact, the degrees of freedom of molecular vibrations can be included in the framework of macroscopic QED, and the relevant studies can be found in our previous works \cite{wang2019quantum,wang2020theory,wang2021simple,wang2022macroscopic}. Next, to derive the polaritonic HR factor, we project Eq.~\ref{Eq:mol_pol_H} into the electronic subspace (see SI section 3) and apply the polaron-type unitary transformation via (see SI section 4)
\begin{align}
\label{Eq:U_photonic}
    \hat{U}_\mathrm{pol} \equiv & \,
    \mathrm{exp}\bigg\{ \int d\mathbf{r}\int_0^\infty d\omega\big(\mathbf{g}_\mathrm{e}(\mathbf{r}_\mathrm{M},\mathbf{r},\omega)\ket{\mathrm{e}}\bra{\mathrm{e}}
    \nonumber \\
    &\hspace{1cm}+\mathbf{g}_\mathrm{\mathrm{g}}(\mathbf{r}_\mathrm{M},\mathbf{r},\omega)\ket{\mathrm{\mathrm{g}}}\bra{\mathrm{\mathrm{g}}}\big)\cdot\hat{\mathbf{f}}(\mathbf{r},\omega)-\mathrm{h.c.}\bigg\},
\end{align}
where $\mathbf{g}_\mathrm{e}(\mathbf{r}_\mathrm{M},\mathbf{r},\omega)$ and $\mathbf{g}_\mathrm{g}(\mathbf{r}_\mathrm{M},\mathbf{r},\omega)$ are defined as
\begin{align}
\label{Eq:gDA_displacements}
    &\mathbf{g}_{\mathrm{e}}(\mathbf{r}_\mathrm{M},\mathbf{r},\omega)=\frac{i\omega}{c^2}\sqrt{\frac{\mathrm{Im}\epsilon_\mathrm{r}(\mathbf{r},\omega)}{\hbar\pi\epsilon_0}}\boldsymbol{\mu}_\mathrm{e}\cdot\overline{\overline{\mathbf{G}}}(\mathbf{r}_\mathrm{M},\mathbf{r},\omega),\\
\label{Eq:gD+A-_displacements}
    &\mathbf{g}_{{\mathrm{g}}}(\mathbf{r}_\mathrm{M},\mathbf{r},\omega)=\frac{i\omega}{c^2}\sqrt{\frac{\mathrm{Im}\epsilon_\mathrm{r}(\mathbf{r},\omega)}{\hbar\pi\epsilon_0}}\boldsymbol{\mu}_{{\mathrm{g}}}\cdot\overline{\overline{\mathbf{G}}}(\mathbf{r}_\mathrm{M},\mathbf{r},\omega).
\end{align}
Here, $\epsilon_0$ is the vacuum permittivity, $c$ is the speed of light in vacuum, $\mathbf{r}_\mathrm{M}$ is the position of the molecular center and $\mathrm{Im}\epsilon_\mathrm{r}(\mathbf{r},\omega)$ is the imaginary part of the complex dielectric function of the environment. $\boldsymbol{\mu}_\mathrm{e}$ and $\boldsymbol{\mu}_\mathrm{g}$ represent the permanent dipole of the excited and ground states, respectively. The dyadic Green's function $\overline{\overline{\mathbf{G}}}(\mathbf{r}_\mathrm{M},\mathbf{r},\omega)$ follows Maxwell's equations $\left(\frac{\omega^2}{c^2}\epsilon_\mathrm{r}(\mathbf{r}_\mathrm{M},\omega) - \nabla \times \nabla \times \right)\overline{\overline{\mathbf{G}}}(\mathbf{r}_\mathrm{M},\mathbf{r},\omega)=-\mathbf{\overline{\overline{I}}}_3\delta(\mathbf{r}_\mathrm{M}-\mathbf{r})$, where $\mathbf{\overline{\overline{I}}}_3$ is a $3\times3$ identity matrix and $\delta(\mathbf{r}_\mathrm{M}-\mathbf{r})$ is the three-dimensional delta function. 

The physical meanings of Eqs.~\ref{Eq:U_photonic}-\ref{Eq:gD+A-_displacements} are elaborated as follows. First, $\mathbf{g}_\mathrm{e}(\mathbf{r}_\mathrm{M},\mathbf{r},\omega)$ and $\mathbf{g}_\mathrm{g}(\mathbf{r}_\mathrm{M},\mathbf{r},\omega)$ indicate that EM polaritonic mode density with the frequency $\omega$ monitored at the position $\mathbf{r}$ interacts with the permanent dipole (electronic displacements) at the position $\mathbf{r}_\mathrm{M}$, leading to a polaritonic displacement per volume per frequency, i.e., unit polaritonic displacement. Next, the spatial and frequency integrals in Eq.~\ref{Eq:U_photonic} sum over all possible unit polaritonic displacements, giving rise to the total polaritonic displacement (dimensionless quantity) which acts at $\mathbf{r}_\mathrm{M}$. As a result, the unitary transformation via Eq.~\ref{Eq:U_photonic} indicates that excited and ground states are shifted along polaritonic coordinates with displacements weighted by local polaritonic density of states ($\mathrm{Im}\overline{\overline{\mathbf{G}}}(\mathbf{r}_\mathrm{M},\mathbf{r}_\mathrm{M},\omega)$) and their corresponding permanent dipoles. 
 
 After the unitary transformation, the transformed electronic and polariotnic Hamiltonian are unchanged (see SI section 4), but the transformed light-matter interaction includes not only the effect of the linear coupling $\boldsymbol{\mu}_\mathrm{eg}\cdot\hat{\mathbf{E}}(\mathbf{r}_\mathrm{M})$ but also the effect of the square term of the polaritonic displacements $\Lambda_\mathrm{pol}$
 as follows,
 \begin{align}
\label{Eq:H_mol_pol_transformed}
    &\hat{\mathscr{H}}_\mathrm{mol\scalebox{0.7}{-}pol}=(-\boldsymbol{\mu}_\mathrm{eg}\cdot\hat{\mathbf{E}}(\mathbf{r}_\mathrm{M})+\Lambda_\mathrm{pol})\ket{\mathrm{e}}\bra{\mathrm{g}}\hat{D}_\mathrm{pol}+\mathrm{h.c.},
\end{align}
where $\boldsymbol{\mu}_\mathrm{eg}$ represents the transition dipole, $\hat{D}_\mathrm{pol}$ stands for the polaritonic displacement operator $\hat{D}_\mathrm{pol}\equiv\mathrm{exp}\bigg\{-\int d\mathbf{r}\int_0^\infty d\omega\Delta\mathbf{g}(\mathbf{r}_\mathrm{M},\mathbf{r},\omega)\cdot\hat{\mathbf{f}}(\mathbf{r},\omega)+\mathrm{h.c.}\bigg\}$ (where $\Delta\mathbf{g}(\mathbf{r}_\mathrm{M},\mathbf{r},\omega)=\mathbf{g}_\mathrm{\mathrm{g}}(\mathbf{r}_\mathrm{M},\mathbf{r},\omega)-\mathbf{g}_\mathrm{e}(\mathbf{r}_\mathrm{M},\mathbf{r},\omega)$). Analogy to reorganization energy due to vibronic displacement, $\Lambda_\mathrm{pol}$ can be regarded as reorganization dipole self-coupling due to polaritonic displacements, and it can be expressed as
\begin{align}
\label{Eq:reorganization dipole-dipole interaction_int}
    \Lambda_\mathrm{pol}=\frac{1}{\pi\epsilon_0 c^2}\int_0^\infty d\omega~\omega\Delta\boldsymbol{\mu}\cdot\mathrm{Im}\overline{\overline{\mathbf{G}}}(\mathbf{r}_\mathrm{M},\mathbf{r}_\mathrm{M},\omega)\cdot\boldsymbol{\mu}_\mathrm{eg},
\end{align}
where $\Delta\boldsymbol{\mu}=\boldsymbol{\mu}_\mathrm{g}-\boldsymbol{\mu}_\mathrm{e}$ is the permanent dipole difference. Notably, the form of Eq.~\ref{Eq:H_mol_pol_transformed} is consistent not only with the light-matter coupling terms derived by Semenov and Nitzan considering single photonic mode \cite{semenov2019electron}, but also with the ones developed by us considering infinite photonic modes\cite{wei2022cavity}. In addition, according to the integral relation of $\mathrm{Im}\overline{\overline{\mathbf{G}}}(\mathbf{r}_\mathrm{M},\mathbf{r}_\mathrm{M},\omega)$ (Eq.~S25 in SI),\cite{buhmann2004casimir} Eq.~\ref{Eq:reorganization dipole-dipole interaction_int} can be reduced to the form of dipole self-energy reported in the previous study \cite{semenov2019electron}, i.e.,
\begin{align}
\label{Eq:reorganization dipole-dipole interaction_Veff}
    \Lambda_\mathrm{pol}=\frac{\Delta\boldsymbol{\mu}\cdot\boldsymbol{\mu}_\mathrm{eg}}{2\epsilon_0 V_\mathrm{eff}}.
\end{align}
Equation~\ref{Eq:reorganization dipole-dipole interaction_Veff} shows that $\Lambda_\mathrm{pol}$ can be viewed as dipole self-energy between permanent dipoles and a transition dipole. By projecting $\hat{D}_\mathrm{pol}$ in the polaritonic subspace, we can define the polaritonic HR factor $S_\mathrm{pol}$ as (see SI section 5)
\begin{align}
\label{Eq:photonic_HR}
S_\mathrm{pol}\equiv\frac{1}{\hbar c^2\pi\epsilon_0}\int_0^\infty d\omega\Delta\mathbf{\mu}\cdot\mathrm{Im}\overline{\overline{\mathbf{G}}}(\mathbf{r}_\mathrm{M},\mathbf{r}_\mathrm{M},\omega)\cdot\Delta\mathbf{\mu}.
\end{align}

\begin{figure}
    \includegraphics[width=\textwidth]{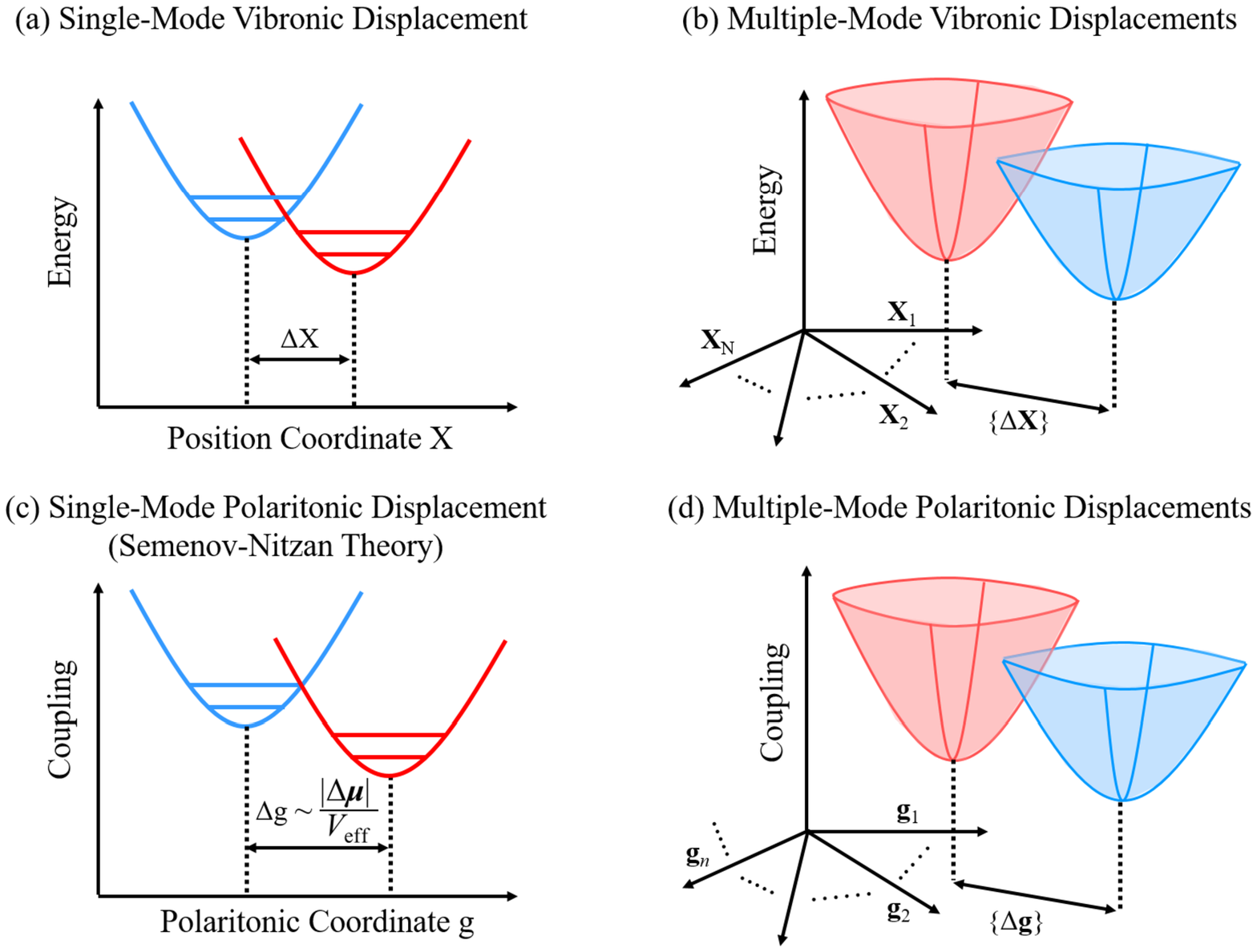}
    \caption{Schematic Illustration of vibronic and polaritonic displacements. (a) Single-mode vibronic displacement. (b) Multiple-mode vibronic displacement with finite discrete normal modes (total number of modes: N). (c) Single-mode polaritonic displacement suggested by Semenov and Nitzan \cite{semenov2019electron}. (d) Multiple-mode polaritonic displacements with infinite continuous modes. Note that the index $n$ represents the $n^\mathrm{th}$ polaritonic coordinate, indicating that the EM polaritonic mode density possesses the $n^\mathrm{th}$ frequency monitored at the $n^\mathrm{th}$ position, i.e., $\Delta\mathbf{g}(\mathbf{r}_\mathrm{M},\mathbf{r}_n,\omega_n) = (\boldsymbol{\mu}_g-\boldsymbol{\mu}_e)\cdot\mathbf{g}_n$, where $\mathbf{g}_n\equiv i\omega/c^2\sqrt{\mathrm{Im}\epsilon_\mathrm{r}(\mathbf{r}_n,\omega_n)/\hbar\pi\epsilon_0}~\overline{\overline{\mathbf{G}}}(\mathbf{r}_\mathrm{M},\mathbf{r}_n,\omega_n)$.}
    \label{Fig:1}
\end{figure}

To clearly elaborate the concept of $\Lambda_\mathrm{pol}$ and $S_\mathrm{pol}$, we compare these two factors with their analogy to traditional Huang-Rhys factors and reorganization energies caused by electron-phonon (or vibronic) couplings in Figure~\ref{Fig:1} and Table~\ref{Table:1}. In addition, Figure~\ref{Fig:1} and Table~\ref{Table:1} emphasize that the concepts of $\Lambda_\mathrm{pol}$ and $S_\mathrm{pol}$ caused by a single mode and by infinite modes are quite different. Figure.~\ref{Fig:1}a illustrates that a single-mode vibronic displacement $\Delta X$ arises from different nuclear configurations of two states, which can be described by a Huang-Rhys factor $S_\mathrm{vib}$. In addition, the square term of $\Delta X$ leads to the energy shift of the effective energy gap, so-called reorganization energy $\lambda_\mathrm{vib}$. The concept of the single-mode Huang-Rhys factor can be generalized to a multiple-mode case, where the total Huang-Rhys factor is associated with the combination of the multiple-mode vibronic displacements along their own normal modes, as shown in Figure~\ref{Fig:1}b \cite{nitzan2006chemical}. By contrast, polaritonic displacements (Figure~\ref{Fig:1}c and \ref{Fig:1}d) can be viewed as the electronic states displaced by light-matter interaction induced by permanent dipoles and local photonic density of states (Eqs.~\ref{Eq:gDA_displacements} and \ref{Eq:gD+A-_displacements}). In our proposed theory, we describe the continuous infinite polaritonic modes via the bosonic vector fields based on macroscopic QED, which leads to multiple unit polaritonic displacements illustrated by $\{\Delta \mathbf{g}\}$ in Figure~\ref{Fig:1}d. This approach allows us to quantitatively describe dipole self-energy and light-matter coupling from permanent dipoles in an arbitrary dielectric environment without free parameters. Notably, if we choose the local polaritonic density of states as a single photonic mode in the long-wavelength approximation (see SI section 6), $S_\mathrm{pol}$ and $\Lambda_\mathrm{pol}$ can be reduced to the single-mode polaritonic displacement developed by by Semenov and Nitzan (Table 1) \cite{semenov2019electron}.

The reorganization dipole self-coupling $\Lambda_\mathrm{pol}$ and the reorganization energy $\lambda_\mathrm{vib}$ play completely different roles in chemistry although the polaritonic Huang-Rhys factor $S_\mathrm{pol}$ and the vibronic Huang-Rhys $S_\mathrm{vib}$ share a similar physical picture. The reorganization dipole self-coupling $\Lambda_\mathrm{pol}$ corresponds to the modification of the coupling strength between two electronic states, while the reorganization energy $\lambda_\mathrm{vib}$ corresponds to the relative energy shift of two electronic states. Recall that Eq.~\ref{Eq:H_mol_pol_transformed} shows that the polaron-like unitary transformation (Eq.~\ref{Eq:U_photonic}) leads to an off-diagonal element ($\Lambda_\mathrm{pol}$) in the electronic subspace. As a result, we emphasize that the y axes in Figure~\ref{Fig:1}c and \ref{Fig:1}d correspond to the coupling strength, not energy.

\clearpage

\renewcommand{\arraystretch}{2}
\begin{table}[!ht]
\footnotesize
\caption{Comparison of vibronic and polaritonic displacements. $V_\mathrm{eff,s}$, $\boldsymbol{\epsilon}_\mathrm{s}$ and $\omega_\mathrm{s}$ indicate the effective mode volume, unit vector and frequency of a single photonic mode, respectively. On the other hand, $V_\mathrm{eff}$ represents the effective mode volume of the infinite polaritonic modes.}
\begin{tabular}{c|c|c} 
\diagbox[width=12em]{Mode Numbers}{Displacements} & Vibronic & Polaritonic \\
\hline
\multirow{2}{4em}{Single} & $S_\mathrm{vib}=\frac{m\omega_\mathrm{vib}\Delta X^2}{2\hbar}$ & $S_\mathrm{pol}=\frac{|\Delta\boldsymbol{\mu}\cdot\boldsymbol{\epsilon}_\mathrm{s}|^2}{2\varepsilon_0\hbar\omega_\mathrm{s} V_\mathrm{eff,s}}$ \\ 
& $\lambda_\mathrm{vib}=\hbar\omega_{\mathrm{vib}}S_{\mathrm{vib}}$ & $\Lambda_\mathrm{pol}=\frac{(\boldsymbol{\mu}_{eg}\cdot\boldsymbol{\epsilon}_\mathrm{s})(\Delta\boldsymbol{\mu}\cdot\boldsymbol{\epsilon}_\mathrm{s})}{2\varepsilon_0V_\mathrm{eff,s}}$ \\ 
\hline
\multirow{3}{4em}{Multiple} & $S_{\mathrm{vib},u}=\frac{m\omega_{\mathrm{vib},u}\Delta X_u^2}{2\hbar}
$ & $S_\mathrm{pol}=\frac{1}{\hbar c^2\pi\epsilon_0}\int_0^\infty d\omega'\Delta\boldsymbol{\mu}\cdot\mathrm{Im}\overline{\overline{\mathbf{G}}}(\mathbf{r}_\mathrm{M},\mathbf{r}_\mathrm{M},\omega')\cdot\Delta\boldsymbol{\mu}$ \\ 
& $S_\mathrm{vib}=\sum_u S_{\mathrm{vib},u}
$ & $\Lambda_\mathrm{pol}=\frac{1}{\pi\epsilon_0 c^2}\int_0^\infty d\omega~\omega\Delta\boldsymbol{\mu}\cdot\mathrm{Im}\overline{\overline{\mathbf{G}}}(\mathbf{r}_\mathrm{M},\mathbf{r}_\mathrm{M},\omega)\cdot\boldsymbol{\mu}_\mathrm{eg}$ \\ 
& $\lambda_\mathrm{vib}=\sum_u \hbar\omega_{\mathrm{vib},u}S_{\mathrm{vib},u}$ & $=\frac{\Delta\boldsymbol{\mu}\cdot\boldsymbol{\mu}_\mathrm{eg}}{2\varepsilon_0 V_\mathrm{eff}}$\hspace{4.45cm} \\ 
\end{tabular}
\label{Table:1}
\end{table}
\renewcommand{\arraystretch}{1}

To investigate the effects of polaritonic displacements on light-matter coupling strength, we apply the Wigner-Weisskopf theory and derive the modified light-matter coupling strength under the resonance condition (see SI section 7).
\begin{align}
    \label{Eq:g}
    g&=\mathrm{exp}\big\{-S_\mathrm{pol}/2\big\}\sqrt{\Lambda_\mathrm{pol}^2+g_0^2},
\end{align}
where $g_0$ corresponds to the light-matter coupling strength caused by only the transition dipole,
\begin{align}
    \label{Eq:g0}
    g_0\equiv\bigg\{\frac{\hbar}{c^2\pi\epsilon_0}\int_0^\infty d\omega~\omega^2\boldsymbol{\mu}_\mathrm{eg}\cdot\mathrm{Im}\overline{\overline{\mathbf{G}}}(\mathbf{r}_\mathrm{M},\mathbf{r}_\mathrm{M},\omega)\cdot\boldsymbol{\mu}_\mathrm{eg}\bigg\}^{1/2},
\end{align}
which has been reported in the previous studies \cite{gonzalez2014reversible,medina2021few,chuang2022tavis}. Notably, as $\Delta\boldsymbol{\mu}=0$, Eq.~\ref{Eq:g} can be reduced to the coupling strength defined in our previous works \cite{wang2019quantum,wang2020theory,wang2021simple}. In addition, Eq.~\ref{Eq:g0} has been used to estimate experimental light-matter coupling strength, which shows good agreement with the experimentally observed Rabi splitting \cite{wang2019quantum,wang2020theory,wang2021simple}.

\begin{figure}
    \includegraphics[width=\textwidth]{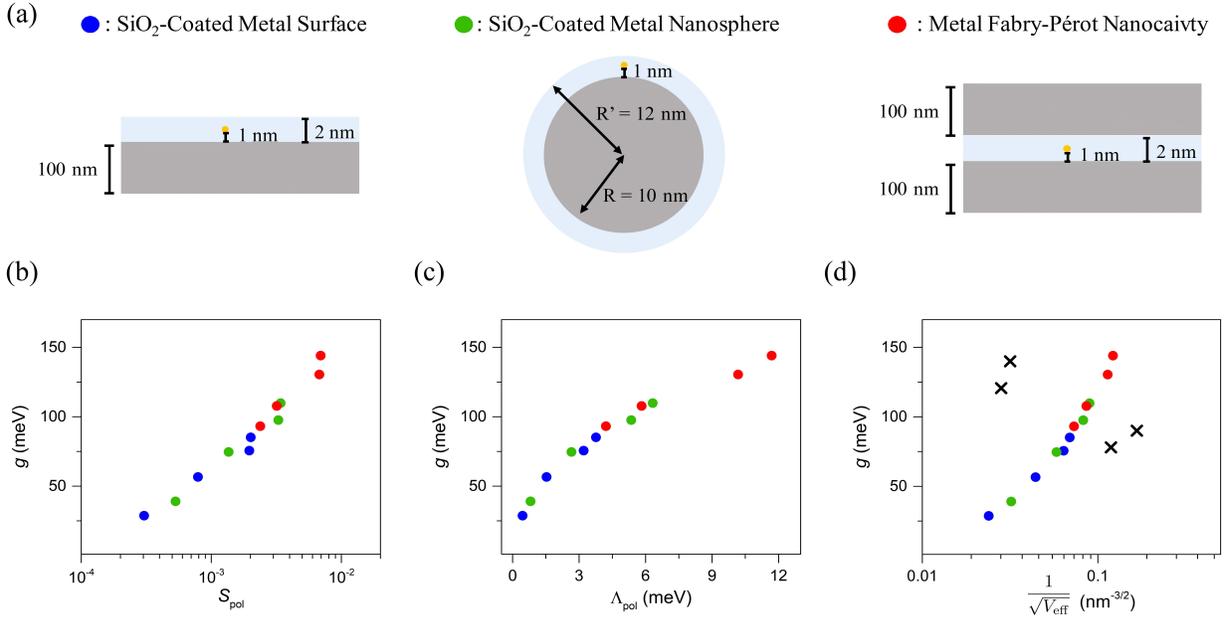}
    \caption{Evaluation of modified light-matter coupling strength ($g$), polaritonic HR factor ($S_\mathrm{pol}$), reorganization dipole self-coupling ($\Lambda_\mathrm{pol}$) and effective mode volume ($V_\mathrm{eff}$). (a) Schematic illustrations of molecules in different dielectric environments. The gray area  represent the metals including Ag, Au, Al and Pt. The light blue area represents the dielectric SiO$_2$. Yellow dots indicate the position of the molecule. The directions of $\Delta\boldsymbol{\mu}$ and $\boldsymbol{\mu}_\mathrm{eg}$ are both perpendicular to the metal surface. The thickness of SiO$_2$ coating of metal surface and nanosphere is 2 nm. (b) Coupling strength vs polaritonic HR factor, (c) Coupling strength vs  reorganization dipole self-coupling and (d) Coupling strength vs inverse root (effective mode volume) for different dielectric environments in (a). The labels "$\cross$" in (d) represent the reported experimental results (Table S2) \cite{chikkaraddy2016single,liu2017strong,park2019tip,li2022room}, showing that the theoretical and the previous experimental results are within three times. The following parameters were used in (b)-(d): |$\boldsymbol{\mu}_\mathrm{eg}| = 10$ Debye and |$\Delta\boldsymbol{\mu}|= 20$ Debye.}
    \label{Fig:2}
\end{figure}

Based on macroscopic QED, our theory not only provides several useful concepts derived from the polaritonic HR factor $S_\mathrm{pol}$, but also offers a quantitative analysis of light-matter interaction in medium. In Figure~\ref{Fig:2}, we evaluate $S_\mathrm{pol}$, reorganization dipole self-coupling $\Lambda_\mathrm{pol}$, modified light-matter coupling strength $g$ and effective mode volume $V_\mathrm{eff}$ in different dielectric environments and compare the reported light-matter coupling strength in experiments.\cite{chikkaraddy2016single,liu2017strong,park2019tip,li2022room}Here we investigate three kinds of photonic systems: SiO$_2$-coated metal surfaces, SiO$_2$-coated metal nanospheres and metal Fabry-Perot (FP) nanocavities, where the metal species include Au, Ag, Al and Pt (Figure~\ref{Fig:2}a). Similar photonic systems showing strong light-matter interaction have been developed in experiments \cite{chikkaraddy2016single,liu2017strong,park2019tip,li2022room,hugall2018plasmonic}.  Here we use the Fresnel method and Mie theory developed by our previous studies to evaluate the dyadic Green's functions in the planar and spherical systems, respectively \cite{wang2020coherent,wang2020theory,lee2020controllable}. The simulated data of each system are shown in Table S1. The numerical details of how to evaluate $g$, $S_\mathrm{pol}$, $\Lambda_\mathrm{pol}$ and $V_\mathrm{eff}$ are shown in SI section 8. Notably, our method is not limited to analytically solvable systems. Through calculating Purcell factor via computational electrodynamics packages, e.g., Lumerical FDTD Solutions \cite{FDTD}, we can numerically obtain a dyadic Green's function in an arbitrary dielectric system\cite{zhou2013lasing,ropp2013nanoscale}.

 The key points of Figure~\ref{Fig:2}b--\ref{Fig:2}d are summarized as follows. First, Figure~\ref{Fig:2}b shows that $S_\mathrm{pol}$ are extremely small ($< 0.01$) even in the FP nanocavity with highly confined EM fields, revealing that it is difficult to create significant polaritonic displacements, e.g., $S_\mathrm{pol} \geq 0.5$, from vacuum fluctuations in most photonic systems. Second, Figure~\ref{Fig:2}c indicates that the correction of light-matter coupling strength from dipole self-energy and permanent dipoles is negligible ($g/g_0 \approx 99.9 \%$) since $\Lambda_\mathrm{pol} \ll g$ and $S_\mathrm{pol} \ll 1$. The decrease of $g$ mainly results from the polaritonic Franck-Condon overlaps (the exponential term in Eq.~\ref{Eq:g}) since the exponential term decays much significantly than the increase contributed by $\Lambda_\mathrm{pol}$.  In addition, even considering different alignments between $\Delta\boldsymbol{\mu}$ and $\boldsymbol{\mu}_\mathrm{eg}$, the ratio $g/g_0$ is only reduced up to 99.7\% when $|\boldsymbol{\mu}_\mathrm{eg}| = 10$ Debye and $|\Delta\boldsymbol{\mu}|= 20$ Debye (Figure~S1). According to Figure~S1, the minimum $g/g_0$ arises as $\Lambda_\mathrm{pol}=0$ owing to the orthogonality of $\Delta\boldsymbol{\mu}$, $\boldsymbol{\mu}_\mathrm{eg}$ and local dyadic Green's function, i.e., $\Delta\boldsymbol{\mu}\cdot\mathrm{Im}\overline{\overline{\mathbf{G}}}(\mathbf{r}_\mathrm{M},\mathbf{r}_\mathrm{M},\omega)\cdot\boldsymbol{\mu}_\mathrm{eg}=0$. Viewed from another perspective, parallel $\Delta\boldsymbol{\mu}$ and $\boldsymbol{\mu}_\mathrm{eg}$ suppress light-matter decoupling and thus promote significant light-matter interaction. Third, Figure~\ref{Fig:2}d demonstrates that the theoretical $g$ and $V_\mathrm{eff}$ (solid circles) and the reported experimental ones (crosses) has the similar order of magnitude, which  indicates that the magnitude of $\Lambda_\mathrm{pol}$ and $S_\mathrm{pol}$ based on our calculation are reliable. Moreover, the link between Figure~\ref{Fig:2}b and \ref{Fig:2}d also shows the concept of the polaritonic HR factor is useful to understand light-matter interaction. Note that the deviation between theoretical and experimental results ($g$ and $V_\mathrm{eff}$) originates from different local polaritonic density of states ($\mathrm{Im}\overline{\overline{\mathbf{G}}}(\mathbf{r}_\mathrm{M},\mathbf{r}_\mathrm{M},\omega)$) and $\boldsymbol{\mu}_\mathrm{eg}$ \cite{chikkaraddy2016single,liu2017strong,park2019tip,li2022room}. The results reveal that our theory can be utilized to predict $g$ and $V_\mathrm{eff}$ in arbitrary dielectric environments, providing a useful guide for experimentalists to precisely control light-matter interaction and QED effects.

\clearpage

\begin{figure}
    \includegraphics[width=0.5\textwidth]{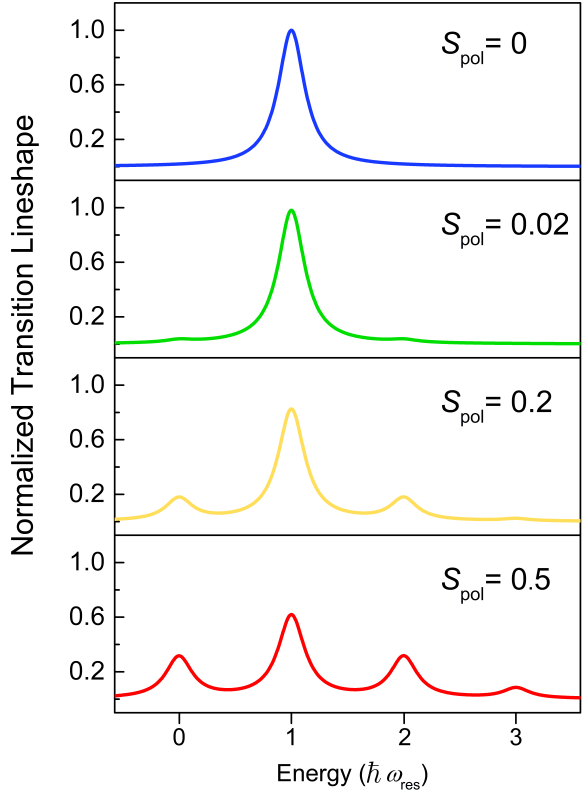}
    \caption{Schematic illustration of polaritonic progression. }
    \label{Fig:3}
\end{figure}
The continuous infinite polaritonic modes can be approximated as an effective polaritonic mode when the spectrum of local polaritonic density of state behaves like single Lorentzian function with resonance frequency $\omega_\mathrm{res}$ \cite{wang2021simple,gonzalez2014reversible,mascherpa2020optimized}.
To explore possible physical phenomena caused by the effective mode with large polaritonic displacements, we illustrate transition lineshapes from the polaritonic ground state with the molecular excited state to all possible single polaritonic states with the molecular ground state under different $S_\mathrm{pol}$ according to the polaritonic Franck-Condon overlaps derived in section 6 in SI (Figure~\ref{Fig:3}). Note that the effective mode contributed by the infinite polaritonic modes cause the linewidths of the transition peaks. For $S_\mathrm{pol}< 0.02$, the transition is mainly driven by light-matter coupling strength $\boldsymbol{\mu}_\mathrm{eg}\cdot\hat{\mathbf{E}}(\mathbf{r}_\mathrm{M})$, leading to the dominance of 0-1 transition. By contrast, for $S_\mathrm{pol}\geq 0.2$, the intensity of 0-1 transition decreases and 0-0, 0-2 and 0-3 transitions becomes significant because polaritonic displacements break the orthogonality of different ladders of polaritonic states. Here we defined the characteristic transition lineshape as "polaritonic progression" analogized by vibronic progression, where non-zero vibronic HR factors cause the vibronic transitions of different vibrational quanta.

The phenomena under large $S_\mathrm{pol}$ are summarized as follows. First, the decrease of 0-1 transition is associated with light-matter decoupling because the definition of light-matter coupling is limited to the coupling of 0-1 transition. Incidentally, the decrease of $g$ is associated with the effects of light-matter decoupling \cite{de2014light,frisk2019ultrastrong}. Second, the rise of 0-2 and 0-3 transitions corresponds to the multipolariton generation, indicating the existence of two-photon and three photon emission processes. Third, the 0-0 transition here represents the "non-radiative" transition induced by light-matter interaction. Our results by considering dipole self-energy and permanent dipole-induced light-matter coupling are consistent with previous reports which suggest the same phenomena caused by dipole self-energy under ultrastrong and deep strong coupling regimes \cite{de2014light,frisk2019ultrastrong,mueller2020deep}. As a result, our theory can also be utilized to predict and simulate light-matter interactions under ultrastrong and deep strong coupling regimes induced by fluctuation EM fields (without applying external fields).

In this study, we have developed several useful concepts based on macroscopic QED and applied them to analyze the effects of dipole self-energy and permanent dipole on light-matter interaction. The main findings are summarized as follows. First, we defined a dimensionless factor to describe how strongly light interacts with matter: polaritonic HR factor $S_\mathrm{pol}$. This factor allows us to quantitatively describe light-matter coupling induced by permanent dipoles and dipole self-energy based on macroscopic QED. Second, inspired by vibronic displacements, we proposed the concept of polaritonic displacement, which enables us to characterize the interactions of infinite polaritonic fields and electronic states and demonstrates the origin of coupling shift dubbed reorganization dipole self-coupling $\Lambda_\mathrm{pol}$. Third, with the aid of macroscopic QED, $S_\mathrm{pol}$, $\Lambda_\mathrm{pol}$ and modified light-matter coupling strength $g$ can be evaluated in arbitrary systems without free parameters, which are in good agreements with the previous experiments. Fourth, we developed the concept of the polaritonic progression and predicted that large $S_\mathrm{pol}$ ($S_\mathrm{pol}>0.2$) gives rise to light-matter decoupling, multipolariton generation and non-radiative transition. Finally, although we do not demonstrate large polaritonic HR factors in the existing photonic systems, it does not mean that large polaritonic HR factors cannot be achieved. For example, the strong polaritonic HR factors may be achieved via collective effects. We believe that our results provide a simple but useful approach for analyzing polaritonic effects, which will stimulate further investigations on the basic principles of QED chemistry and polariton chemistry.

\begin{acknowledgement}

We thank Ming-Wei Lee for the discussion in theoretical derivations. We thank Academia Sinica (AS-CDA-111-M02) and the Ministry
of Science and Technology of Taiwan (MOST 110-2113-M001-053) for generous support.

\end{acknowledgement}

\begin{suppinfo}

The Supporting Information is available free of charge.

G{\"o}oeppert-Mayer transformation (derivation of equation 1); Property of imaginary part of Dyadic Green’s function; model Hamiltonian; Unitary transformation along polaritonic coordinates; Derivation of polaritonic Huang-Rhys factor; Conversion of $S_\mathrm{pol}$ and $\Lambda_\mathrm{pol}$ from macroscopic QED version to cavity QED version; Derivation of modified light-matter coupling strength; Numerical parameters for evaluating $g$, $S_\mathrm{pol}$, $\Lambda_\mathrm{pol}$ and $V_\mathrm{eff}$; Supplementary Figure 1; Supplementary Table 1-2.

\end{suppinfo}

\bibliography{achemso-demo}

\end{document}